\documentstyle[11pt,newpasp,psfig,twoside]{article}
\newcommand{\mum}{$\,\mu$m}

  \def\itm#1 {\vskip10pt \noindent \square\ {\bf #1} }
  \def\square {\hbox{\vrule width5pt height5pt}}

\def\edcomment#1{\iffalse\marginpar{\raggedright\sl#1\/}\else\relax\fi}
\reversemarginpar

\begin{document}
\title{A best fit evolution scenario for the SCUBA galaxies}
 \author{S.\,C.\ Chapman}
\affil{California Institute of Technology,
Pasadena, CA 91125,~~U.S.A.}
\author{G.\,F.\ Lewis}
\affil{Anglo-Australian Observatory, P.O. Box 296, Epping,
        NSW 1710, Australia}
 \author{G.\ Helou}
\affil{IPAC, California Institute of Technology,
Pasadena, CA 91125,~~U.S.A.}

\begin{abstract}
We explore the evolution of the SCUBA sub-mm galaxy population using
Monte Carlo simulations to generate synthetic sub-mm/radio color magnitude
diagrams.
To represent the local distribution of observed dust properties, we use a
local far-infrared luminosity function derived from the IRAS 1.2\,Jy
catalog, bivariate in FIR luminosity and 60\mum/100\mum\ color.
We assume a peak luminosity evolution scenario and a fixed cosmology,
to fit a single parameter model to the existing sub-mm/radio data.
Our best fitting model has a peak in the luminosity evolution at $z=2.6$.
\end{abstract}

\section{Introduction}

Sub-mm luminous, extragalactic
sources are now routinely detected with the SCUBA/JCMT and MAMBO/IRAM 
instruments, and over 100 blank field sources are now
known (Smail et al.~2002, Hughes et al.~1998,
Barger et al.~1999, Chapman et al.~2002a, Eales et al.~2000, Webb et al.~2002,
Borys et al.~2002, Scott et al.~2002, Dannerbauer et al.~2002).
However, the population
continues to be poorly understood, largely as a result of two
observational difficulties. Firstly,
obtaining large samples of objects from sub-mm mapping is a time consuming
process, whereby one source is uncovered in a night worth of
blank sky integration.
Secondly, identifying secure detections at other wavelengths is
difficult due to the positional uncertainty and
large beam sizes ($\sim$15 arcsec),
and the intrinsic faintness of most sources at all other wavelengths.

The tight correlation observed locally between thermal far-IR emission and
synchrotron radio emission (Helou et al.~1985, Condon 1992) suggests a
possibility for identifying the sub-mm sources. The positional accuracy
and small beam sizes of large radio interferometers like the VLA can act as
a surrogate to the sub-mm, allowing precise identifications at other
wavelengths. Smail et al.~(2000) demonstrated that a significant fraction
of their sub-mm sources could be detected in the radio.
Radio surveys, however, do not benefit from the negative K-correction inherent
in sub-mm surveys, and galaxies with luminosities similar to the local ULIG, 
Arp220, will likely be missed by $z\sim3$ in the radio.
Barger, Cowie \& Richards (2000) and Chapman et al.~(2001) have demonstrated that
selecting the optically fainter tail of the microjansky radio sources can 
be used to rapidly uncover a large portion of the blank field sub-mm population
($\sim$70\% of the bright counts). Chapman et al.~(2002b -- hereafter C02) have
presented the results of a large campaign using the radio pre-selection 
technique to uncover sub-mm sources.
 
In this contribution, we continue to investigate
the evolutionary behavior of the sub-mm population of galaxies,
expanding on the models presented in C02. The key addition is an explicit
model of the dust temperature distribution found locally. 
We use these models to better understand the range
in properties sampled by the radio pre-selected sub-mm galaxy population.
Our calculations are done in a $\Lambda=0.7$, $\Omega=0.3$, h=0.65 cosmology.

\section{Fitting the evolution model}
We anchor our model to the local FIR luminosity function (LF), constructed
from the 1.2\,Jy sample of IRAS galaxies (Fisher et al.~1995).
Recent work has emphasized the variation in dust temperature found in
luminous infrared galaxies (Chapman et al.\ 2002c). We therefore form
an LF which represents the distribution in dust temperatures found in
the 1.2\,Jy sample, parametrized by the 60\mum/100\mum\ color
(a full characterization of this LF
will be presented in Chapman et al., in preparation).

We evolve this local bivariate LF
using pure luminosity evolution of the form
$\Phi (L,\nu) = \Phi _0 (L/g(z),\nu (1+z))$.
Our evolution function has a power law peak, $g(z) = (1+z)^{4}$ out
to a break redshift, and drops thereafter as $g(z) = (1+z)^{-4}$.
This power-law index is
chosen based on evolutionary models fit to both optical and sub-mm wavelength
data (Blain et al.~1999a,b).
The 60\mum/100\mum\ color distribution does not evolve, but rather continues
to scale with FIR luminosity as found locally.
The model therefore has only one free parameter, the $g(z)$ break redshift.  
We then adopt a Monte Carlo approach, drawing luminous infrared
galaxies randomly from the evolving distribution function. 
A galaxy thus selected is then assigned a template spectral energy 
distribution from the catalog of Dale et al.~(2001, 2002),
parameterized by the 60\mum/100\mum\ color.
This model scenario does not incorporate the intrinsic scatter in the 
far-IR/radio correlation, as was done in C02. Rather we concentrate on the
properties of the intrinsic dust temperature distribution.
 
In Fig.~1, we plot the distribution of measured sub-mm sources from C02 
in the sub-mm/radio color-magnitude 
diagram (CMD), along with similar CMDs from
our model for three
peak redshifts, $z=2.0$, $2.5$, and $3.0$. Objects which are detectable
in a 30$\mu$Jy radio survey are shown as larger symbols.

We then fit the model galaxy distributions in the CMD to the C02 dataset,
parameterizing only 
by the $z_{\rm break}$ redshift of the peak evolution function.
We minimize the residuals between the real and modeled galaxy distributions
over the range S$_{850}=5-20$\,mJy.
A best fitting evolution is found for a peak redshift of $z=2.6$.

%
% FIGURE (1) --
%
%
\begin{figure*}[htb]
\centerline{
\psfig{file=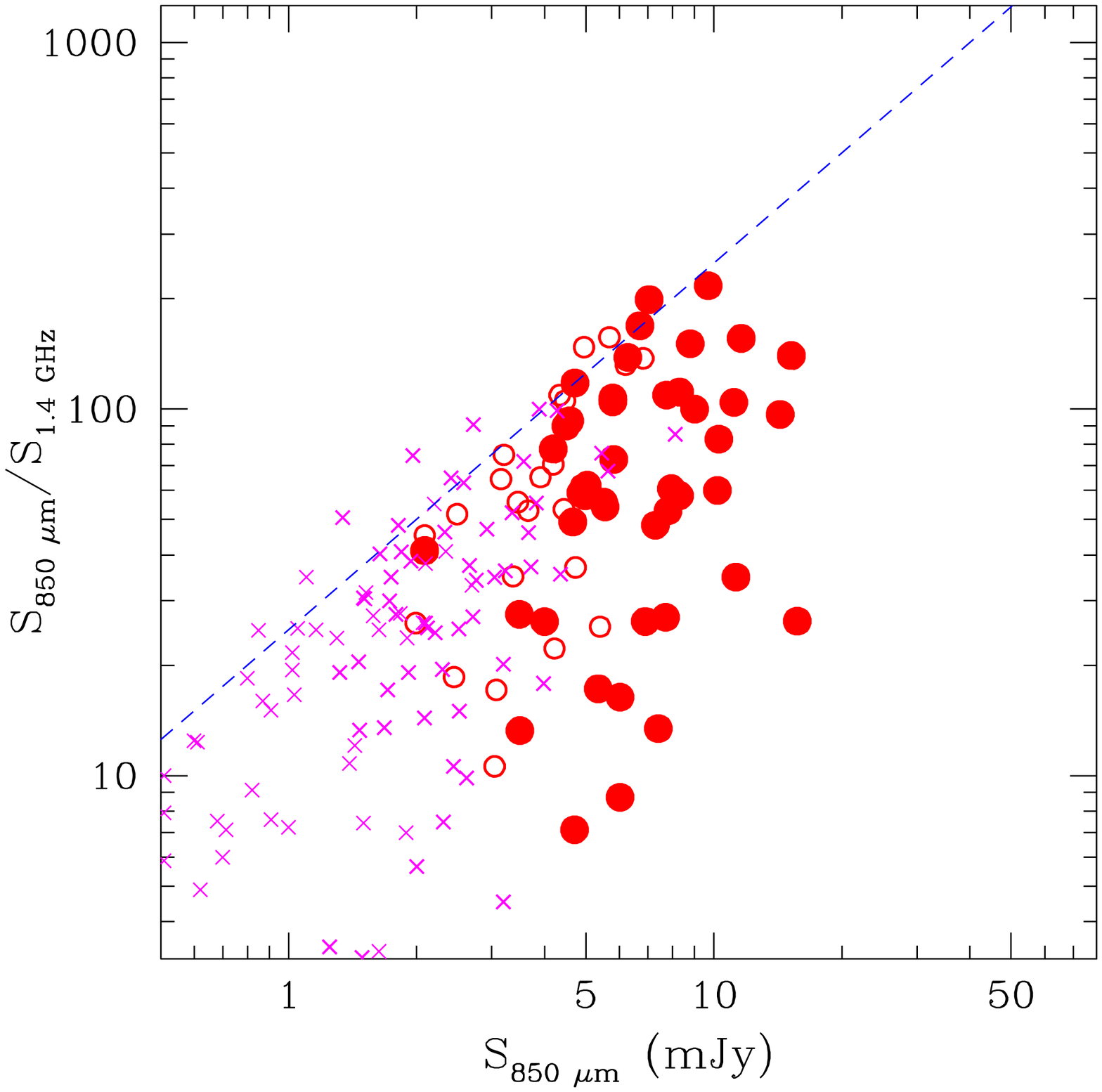,height=6.25cm,angle=0}
\psfig{file=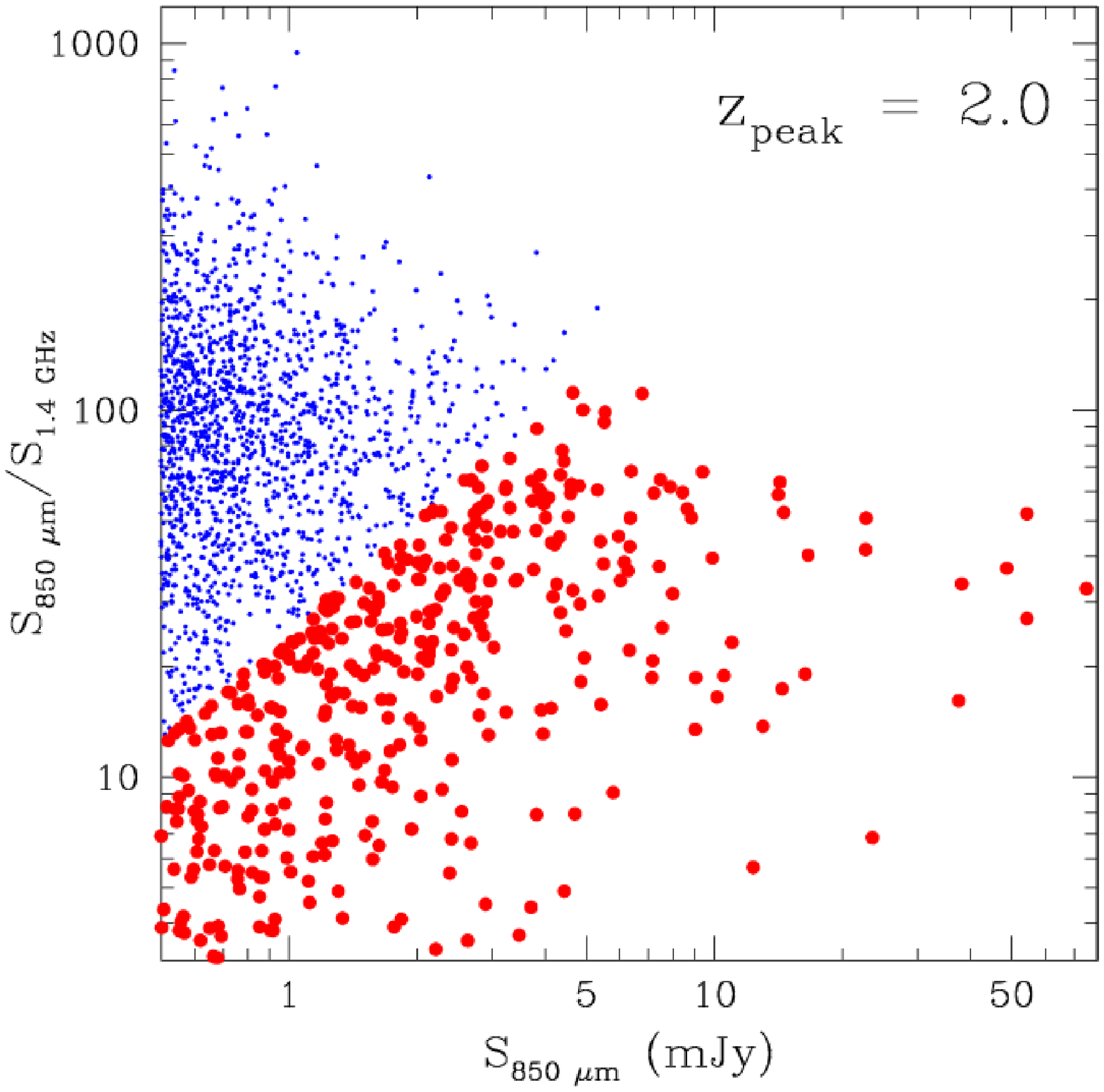,height=6.25cm,angle=0}}
\centerline{
\psfig{file=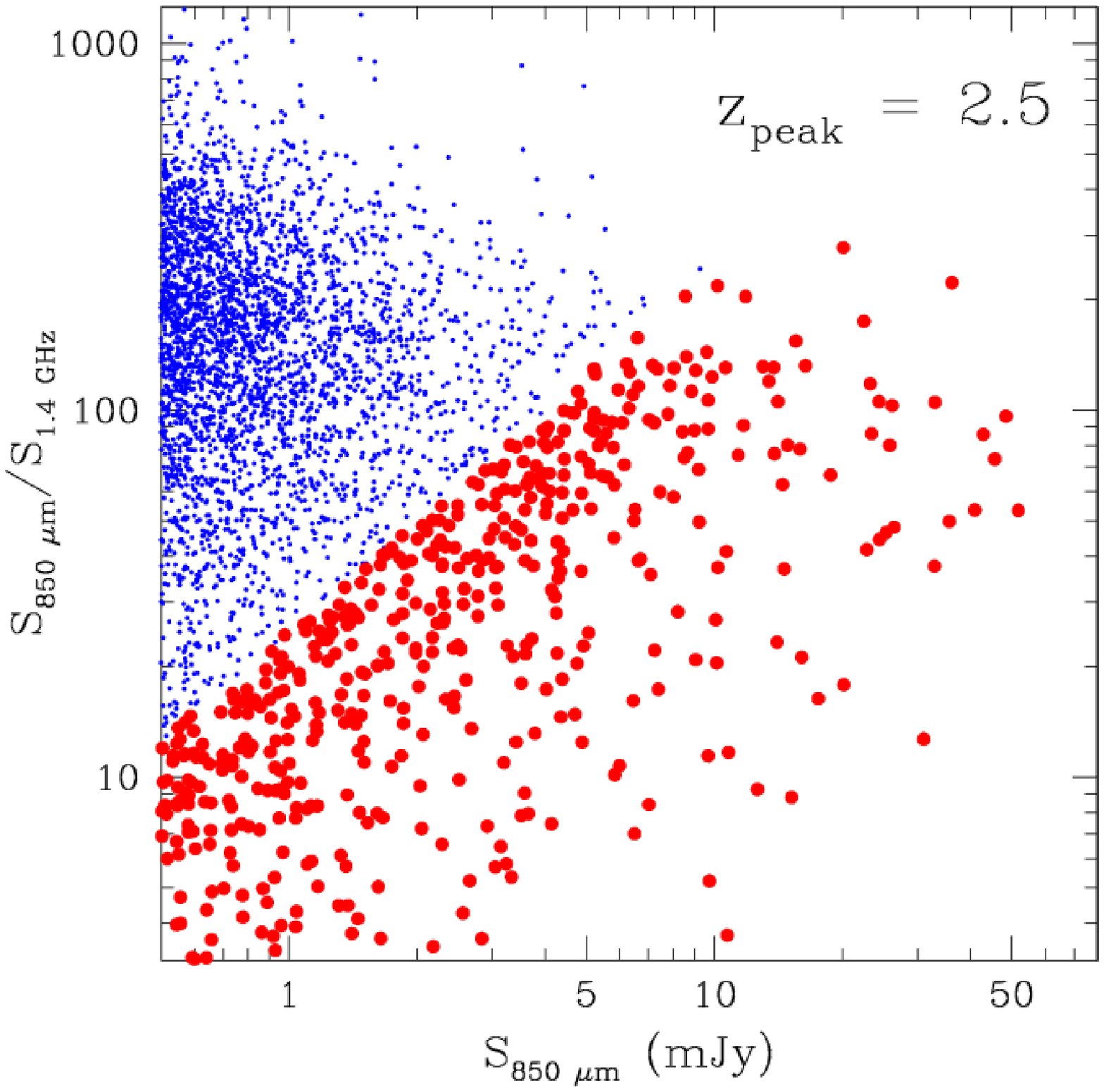,height=6.25cm,angle=0}
\psfig{file=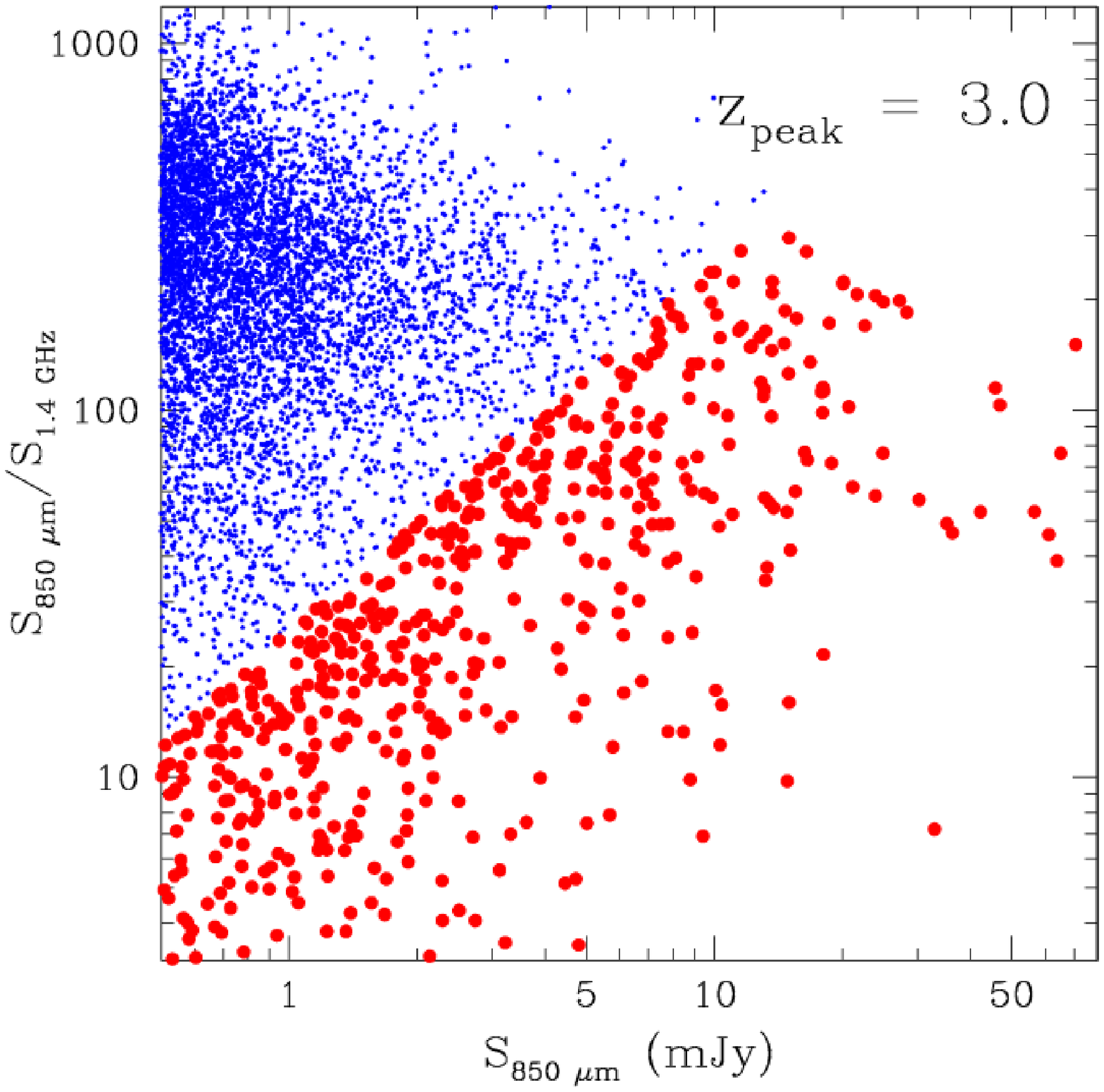,height=6.25cm,angle=0}}
\label{fig1}
\caption{
The S$_{850 \mu m}$, radio color-magnitude diagram from C02, along with three
example model outputs for luminosity evolution peaks at $z=2, 2.5, 3$.
In the data CMD, solid red circles show sub-mm detected sources, open
circles show marginally detected sub-mm sources, and crosses show
radio sources without significant sub-mm detection. The dashed line depicts
the radio limit of the C02 survey.
Model CMDs show larger symbols having 1.4\,GHz flux greater than 40$\mu$Jy.
}
\end{figure*}

\section{Discussion}
The addition of the 60\mum/100\mum\ color distribution to our adopted
FIR luminosity function results in
a fraction of cold luminous galaxies which are preferentially selected
with SCUBA at 850\mum (e.g., Blain 1999, Eales et al.~1999).
As expected (see also the discussion in C02), the peak evolution 
redshift must be 
lower if a population of luminous colder sources exist at high redshifts. 
By comparison,
in C02, a reasonable fit to the data was found for a peak evolution redshift
of $z=3$, when all objects were assumed to be as hot as Arp220 ($\sim50$\,K).

Note also that in this model, there are very few bright sub-mm sources missed
by the 30$\mu$Jy radio threshold. Any model with a peak in the luminosity
evolution function will exhibit similar behavior.
No bright blank field sub-mm 
sources have yet been identified with galaxies
at $z>3$. Our model fit to the existing radio/sub-mm data
suggests a paradigm where this may remain largely 
true even with complete identifications.
As redshifts are obtained for the radio/sub-mm sources, tighter constraints
on the form of the evolution will be possible, breaking the degeneracy of
dust temperature versus redshift. 

\section*{Acknowledgements}
We thank all our collaborators on the various aspects of this project
for a stimulating romp through the enigmatic field of sub-mm galaxies.
SCC gratefully acknowledges support from NASA through
HST grant 9174.1.
GFL thanks the Australian Nuclear
Science and Technology Organization (ANSTO) for financial support.

% REFERENCES
\section*{references}

\noindent  Barger, A., Cowie, L., Richards, E., 2000a, AJ 119, 2092 (BCR)

\noindent  Barger, A., et al., 1999, AJ 117, 2656  %counts

\noindent  Blain, A., et al., 1999a, MNRAS 302, 632 %empirical evol

\noindent  Blain, A., et al., 1999b, MNRAS, 309, 715 %hierarch evol

\noindent  Blain, A., 1999, MNRAS, 309, 955 %dust temp and CY

\noindent  Borys, C., Chapman, S.C., Scott, D., Halpern, M., 2001, 
        MNRAS, in press

\noindent  Chapman, S.C., Richards, E., Lewis, G.F., et al.,
        2001, ApJL, 548, 147 

\noindent  Chapman, S.C., Scott, D., Borys, C., Fahlman, G., 2002a, MNRAS,	
	330, 92 

\noindent  Chapman, S.C., et al. 2002b, ApJ, in press, (astro-ph/0111157) -- C02

\noindent  Chapman, S.C., Smail, I., Ivison, R., et al., 2002c, ApJ, in press, 
	(astro-ph/0203068)

\noindent  Condon, J., 1992, ARA\&A, 30, 575

\noindent  Dale, D., et al., 2001, ApJ, 549, 215

\noindent  Dale, D., et al., 2002, ApJ, in press

\noindent  Dannerbauer, G., et al., 2002, A\&A, in press

\noindent  Eales, S., et al., 2000, AJ, 120, 2244

\noindent  Eales, S., et al., 1999, ApJ 515, 518

\noindent  Fisher, K., et al., 1995, ApJS, 100, 69

\noindent  Helou, G., et al., 1985, ApJ 440, 35

\noindent  Hughes, D., et al., Nature, 1998, 394, 241

\noindent  Scott, S., et al., 2002, MNRAS, in press

\noindent  Smail, I., Ivison, R., Blain, A., Kneib, J.-P., 2002, MNRAS, in press

\noindent  Smail, I., et al., 2000, ApJ 528, 612

\noindent  Webb, T., et al., 2002, ApJ, submitted

\end{document}